# A gamma-ray spectroscopy survey of Omani meteorites

Patrick WEBER[1*], Beda A. HOFMANN[2], Tamer TOLBA[3], Jean-Luc VUILLEUMIER[3]
[1]Hôpital Neuchâtelois, Service de Radiothérapie, Rue du Chasseral 20,
CH-2300 La Chaux-de-Fonds
[2]Naturhistorisches Museum der Burgergemeinde Bern, Bernastrasse 1, CH-3005 Bern, Switzerland
[3]Albert Einstein Center for Fundamental Physics, LHEP, University of Bern, Sidlerstr. 5, CH-3005 Bern, Switzerland
*Corresponding author. E-mail: Patrick.Weber2@h-ne.ch

**Abstract**-The gamma-ray activities of 33 meteorite samples (30 ordinary chondrites, 1 Mars meteorite, 1 iron, 1 howardite) collected during Omani-Swiss meteorite search campaigns 2001-2008 were nondestructively measured using an ultra-low background gamma-ray detector. The results provide several types of information: Potassium and thorium concentrations were found to range within typical values for the meteorite types. Similar mean $^{26}Al$ activities in groups of ordinary chondrites with a) weathering degrees W0-1 and low $^{14}C$ terrestrial age and b) weathering degree W3-4 and high $^{14}C$ terrestrial age are mostly consistent with activities observed in recent falls. The older group shows no significant depletion in $^{26}Al$. Among the least weathered samples one meteorite (SaU 424) was found to contain detectable $^{22}Na$ identifying it as recent fall close to the year 2000. Based on an estimate of the surface area searched, the corresponding fall rate is ~120 events/$10^6$ km$^2$*a, consistent with other estimations. Twelve samples from the large JaH 091 strewn field (total mass ~4.5 t) show significant variations of $^{26}Al$ activities, including the highest values measured, consistent with a meteoroid radius of ~115 cm. Activities of $^{238}U$ daughter elements demonstrate terrestrial contamination with $^{226}Ra$ and possible loss of $^{222}Rn$. Recent contamination with small amounts of $^{137}Cs$ is ubiquitous. We conclude that gamma-ray spectroscopy of a selection of meteorites with low degrees of weathering is particularly useful to detect recent falls among meteorites collected in hot deserts.

## INTRODUCTION

The sultanate of Oman has become an important source for hot desert meteorites since 1999 (Al-Kathiri et al., 2005; Bevan, 2006; Grossman, 2000; Hofmann et al., 2003). The country is gradually becoming the only large hot desert meteorite find area for which a high standard of documentation has been maintained for all finds. It is an ideal area to study the behavior of a hot desert meteorite population under influences of weathering. Fourteen Omani-Swiss meteorite search campaigns (2001-2015) yielded 6041 meteorite samples corresponding to ~880 fall events. More than 100 samples are $^{14}C$-dated (Al-Kathiri et al., 2005; Jull et al., 2008; Zurfluh et al. 2016) demonstrating that terrestrial ages for the Oman meteorite population range from <1 kyr to >50 kyr with a median value close to 20 kyr.

Surveys of meteorite populations using non-destructive gamma-ray spectroscopy (GRS) have proven useful for detection of samples with very high terrestrial ages using $^{26}Al$ ($t_{1/2}$ 717 kyr) in Antarctic meteorites (Evans et al., 1992; Wacker, 1997; Welten et al., 1999) as well as in recent falls to detect short-lived isotopes (e.g. $^{22}Na$, $t_{1/2}$ 2.6 years). We have conducted a survey of 32 meteorite samples recovered from the Oman desert in order to detect possible samples with high (using $^{26}Al$) and very low (live $^{22}Na$) terrestrial age, but also to check for contamination by natural U and Th series isotopes and man-made $^{137}Cs$. Several samples taken from the very large (51 km long) JaH 091 strewn field were measured to test for a possible sorting of samples derived from different depths in the meteoroid within the strewn field.

## SAMPLES AND METHODS

### Samples

The analyzed samples were all collected by Omani-Swiss meteorite search campaigns between 2001 and 2008. All meteorite samples are officially named, or pairings of named meteorites in case of large strewn fields, and locality and find information can be accessed through the database of the Meteoritical Society. All analyzed samples are currently stored in the Natural History Museum Bern (NMBE) as a special collection owned by the Sultanate of Oman. Samples were collected without touching in the field and wrapped in aluminum foil and/or polypropylene bags. In the laboratory, all meteorites were cleaned from loose soil using pressurized air. Cutting to obtain classification samples was done using isopropanol as the only coolant. Many of the samples selected were previously dated using the $^{14}C$ method (Al-Kathiri et al., 2005, Zurfluh et al. 2016). The samples analyzed mostly represent (near)-complete individuals or naturally broken fragments. They typically include a significant amount of naturally weathered (and contaminated) surface. Samples for the GRS survey presented in this work were selected based on the following criteria: i) samples representing a variety of terrestrial ages; ii) least weathered meteorites in order to detect recent falls. This group of samples was carefully selected to represent the least weathered meteorites found based on low degrees of weathering (W0-1); iii) "old" samples with low activities of $^{14}C$ and/or high weathering degrees (W3.6-W4.5, based on refined weathering scale of Zurfluh et al. 2016); iv) a group of samples from the large JaH 091 strewn field in order to sample a large meteorite; v) unusual and uncertain meteorites; vi) soil samples collected close to meteorite find sites. The two soil samples represent the sieved fraction <0.1 mm, collected at the following coordinates: 0101-157: 18° 33.543'N 54° 5.895'E (Dhofar); 0503-128B: 19° 41.586'N 56° 39.263'E (JaH 091 strewn field, location of JaH 090).

The JaH 091 strewn field was shortly described by Gnos et al. (2006) and Zurfluh et al. (2011). This strewn field has a size of 51.2 by 7.2 km. 701 individually fallen masses (most of the largest ones broken up) were collected and geomagnetic surveys at the find locations of the largest masses revealed the presence of unrecovered material, with an estimated mass of 670 kg. Including the paired meteorite JaH 055, the total mass of identified material amounts to ~4500 kg corresponding to a minimum radius of 0.7 m (no ablation) or more likely 0.9-1.2 m (50 and 80% mass loss, respectively).

### Gamma-ray spectroscopy

Gamma-ray measurements of radionuclides were performed with an ultra-low noise germanium detector (Gonin et al. 2003, Leonard et al. 2008) used mostly for material screening for the Enriched Xenon Observatory (EXO) experiment, searching for neutrinoless double beta decay in $^{136}Xe$. The detector itself, made by Eurisys-Mesures, is a 400 $cm^3$ single germanium crystal, p type coaxial. The energy resolution is 1.4 keV FWHM at 238 keV, and 2.5 keV at 1460 keV, scaling with the square root of the energy above that. Low energy tails appear at higher energies. The detector is located in the road tunnel of "La Vue-des-Alpes" (10 km northwest of Neuchâtel, Switzerland), providing a shielding against cosmic rays corresponding to 600 meters water-equivalent, reducing the cosmic muon flux by a factor 1000, and completely suppressing the neutrons. The germanium detector is moreover shielded against the rock radioactivity with an ultraclean lead shielding of 20 cm thickness and a copper shielding of 15 cm thickness. Both copper and lead were found clean of radiocontaminations with a previous germanium detector, which had a sensitivity of $10^{-8}$ g/g, for $^{238}U$ and $^{232}Th$. Finally, the shielding is enclosed in an airtight box and the detector is protected against remnant radon present in the air of the laboratory by a slight nitrogen overpressure replacing the air around the detector. It has been demonstrated that radon contamination is completely flushed out from the measurement volume after 5 hours. Gamma-ray spectroscopy data are only collected 24 hours after sample change to completely eliminate radon contamination from the spectra (Gonin et al. 2003, Leonard et al. 2008). Peak intensities are obtained by integrating the counts in an energy window encompassing the peak. No peak fitting is done. The continuous background under the peak is subtracted. In most cases the background is seen to be rather constant, and is determined in an energy window above the peak. In certain cases it was however necessary to use a linear background.

The spectrum of a typical meteorite is shown in Figure 1. It is compared with the Germanium detector background spectrum. After subtracting this background, depending on the sample size, sensitivities down to $10^{-12}$ g/g can be achieved for $^{238}U$ and $^{232}Th$ natural decay chains, and $10^{-20}$ g/g for short-lived radioisotopes, such as $^{137}Cs$, $^{60}Co$ and $^{22}Na$.

The $^{238}U$ activity is reported first as the mean activity of five daughter element peaks: $^{214}Bi$ (Energy/Branching ratio: 609.3keV/46.1%; 1120.3keV/15.1%; 1764.5keV/15.4%) and $^{214}Pb$ (295.2keV/19.3%; 351.9keV/37.6%), assuming secular equilibrium. It is thus an equivalent activity, labelled $^{238}U_{eq}(214)$. Some caution is necessary however, since these nuclides come after $^{222}Rn$ in the decay chain, which has a long half life of 3.82 days and may diffuse out to a certain extent. The $^{238}U$ daughter $^{226}Ra$ comes before; for this reason, it is treated separately. Also the low-energy peak of this element (186.2 keV, branching ratio of 3.6 %) overlaps with a $^{235}U$ peak (185.7 keV, 57%). The $^{226}Ra$ intensity was calculated under the hypothesis of 0.72 % abundance of $^{235}U$ relative to $^{238}U$ and secular equilibrium. From the $^{226}Ra$ intensity, still assuming secular equilibrium the activity $^{238}U_{eq}(226)$ was derived. A suppression of $^{238}U_{eq}(214)$ compared to $^{238}U_{eq}(226)$ is indicative of Rn emanation. Moreover, an excess can be traced to the special chemical affinity of $^{226}Ra$, which leads to diffusion into the meteorite from the Omani desert soil.

The $^{232}Th$ activity is the mean activity of six daughter element peaks: $^{212}Pb$ (238.6keV/43.3%), $^{228}Ac$ (338.3keV/11.3%; 911.2keV/25.8%; 968.9keV/11.8%) and $^{208}Tl$ (583.2keV/28.1%; 2614.5keV/38.0%), assuming secular equilibrium, and is labeled as $^{232}Th_{eq}$.

$^{40}K$ was also measured and reported as K in the Table 1, assuming a relative concentration of 0.012% of $^{40}K$ in the natural K. It was mostly used to check the measurement calibration, as the concentration of K is almost constant for all meteorite samples.

The following cosmogenic isotopes, mainly activated by Galactic Cosmic Rays (GCR), were measured:

- $^{26}Al$ ($t_{1/2}$ 7.17 $10^5$ years) was measured on one hand to recognize possible meteorites of very high terrestrial ages, and on the other hand, in the case of the JaH-091 strewn field, the $^{26}Al$ activity is an indication of the position (shielding depth) of the meteorite fragment in the meteoroid before atmospheric entry.
- $^{22}Na$ ($t_{1/2}$ 2.6 years) was systematically measured also, in order to scope for low terrestrial age samples.
- Finally, $^{60}Co$ ($t_{1/2}$ 5.27 years) and man-made $^{137}Cs$ ($t_{1/2}$ 30.08 years) were also measured for each sample.

**Detector efficiency calculation**

In order to obtain quantitative values for the activities per unit mass of all the isotopes listed above, a calculation of an energy dependent efficiency factor, namely the probability of full energy deposition in the sensitive volume was performed, using GEANT 4, taking into account:

1. the meteorite sample geometry and chemical composition
2. the self-absorption and the Compton scattering of the emitted gammas in the meteorite, the detector cryostat wall and the germanium dead layer.
3. pile ups in the case of the simultaneous emission of gammas, or of a positron and a gamma; and the production of electron-positron pairs by gammas in the sample or the detector, leading to single and double escape peaks.

In the case of meteorite samples, the chemical compositions and densities are well known, but the sample geometries exhibit irregularities that cannot be fully taken into account.
In the GEANT4 simulation, sample shapes were approximated by cuboids, resulting in a correction factor assumed to be realistic of the sample's self-shielding.
In order to verify this assumption, we measured the gamma activity of two samples from the meteorite Al Huqf 010 (L6 type).
The first sample had a mass of 155.4 g and an irregular shape, scaled to a cuboid with dimensions 5.00 x 3.50 x 2.65 cm in the simulation.
The second sample has been cut to a real cuboid, with dimensions 4.41 x 3.24 x 2.97 cm.
Both samples were gamma counted, the first irregular sample for 5.35 days and the second for 6.34 days. The results are summarized on Table 2.

The results show a very good agreement between the cuboid cut sample and the irregular shaped sample with scaled dimensions. This is true for every of the measured radio-isotopes, except for $^{238}$U. This difference was expected, as $^{238}$U comes from a soil contamination of the sample. It is accumulated close to the samples surface, and the $^{238}$U value is thus lower on the cut sample, where its surface has been removed.

Concerning pile ups, they are important with a detector our size. For gammas, they are particularly important in $^{208}$Tl ($^{232}$Th chain) and $^{214}$Bi ($^{238}$U chain). Coincident gamma rays were generated simultaneously according to the decay scheme. All relevant lines with intensities above 1% were retained. The two gammas in $^{60}$Co were also generated simultaneously. In $^{26}$Al the positron annihilation is followed by the emission of two 511 keV gamma rays, leading to significant pile up effects. Positrons were generated along with the 1808 keV gamma. The 1808 keV full energy peak is clearly seen in the measured spectrum of Figure 1, as well as the sum peak with one of the 511 keV annihilation gammas at 2319 keV. The simulation reproduces correctly the relative intensities of the two peaks. A similar effect is expected for $^{22}$Na. The calculated efficiency for a typical meteorite assuming single gammas is shown in Figure 2, with the efficiency corrected for pile ups in the relevant cases. The effect is seen to be important.

Two different cross-checks of the efficiency calculation were performed, the first one with a sample from the JaH 073 strewn field (No 0201-168), measured independently by AMS for $^{26}$Al (Huber et al. 2008), the second one with 3 samples (Aybut 001, JaH 091 (sample 0603-8), Dho 806), using an independent germanium detector at the Fréjus gamma-spectroscopy laboratory for $^{26}$Al, $^{40}$K, $^{137}$Cs, $^{238}U_{eq}$ and $^{232}Th_{eq}$ activities. The results are listed in Table 3 and show that the measurements in different laboratories yield comparable values. GRS data were obtained by counting for 7 days for every single meteorite sample, discarding the first 24 hours because of radon, as explained above.

Data reduction from GRS assumes a homogeneous distribution of radionuclides within the sample, which is reasonable for K and cosmogenic isotopes considering the size of the samples analyzed (<15 cm). However, it is likely that contaminating isotopes (U- and Th-series as well as $^{137}Cs$) are particularly enriched in the outer zone of the meteorites, as is observed for the main contaminant Sr (Zurfluh et al. 2016), as discussed above. Absolute values for contaminant radionuclides are therefore likely overestimated.

We thus conclude that the Monte-Carlo simulation produces a correct value for the self-shielding correction factor. Conservatively we assign a 10 % systematic error to that factor, which dominates the uncertainty on the determination of the count rates.

## RESULTS AND DISCUSSION

Data of the gamma-ray spectroscopic analysis of 33 meteorite samples (23L, 4H, 3LL, 1 howardite, 1 iron, 1 Martian) and two soil samples are presented in Table 1. In comparison with previous similar surveys (Nishiizumi 1987, Evans 1992, Welten 1999) we provide data for additional radionuclides such as the U and Th series and $^{137}Cs$. Data for $^{40}K$, $^{26}Al$ and $^{22}Na$ can be compared with data from other meteorite surveys. In Table 1 a positive result is given if it is above 3 sigma (99.73 %), else a 90 % CL upper limit is indicated.

### Potassium in ordinary chondrites

The mean measured K activity for ordinary chondrites of 1470 ± 240 dpm/kg corresponds to an average total K concentration of 760 ± 110 ppm, in good agreement with the mean for dominant L- and LL-chondrites of 825/790 ppm (Wasson and Kallemeyn 1988). Old and young age groups show identical mean values of 845 ± 140 and 745 ± 80 ppm, respectively. The two H chondrites yield 700 and 640 ppm. The samples from the JaH 091 strewn field yield a mean value of 710 ± 90 ppm K. Potassium measured in iron meteorite Shişr 043 is likely due to soil particles firmly enclosed in the oxidation rind. Oxide flakes contain up to 0.11% $K_2O$ (Al-Kathiri et al. 2006). Howardite JaH 556 yields a K content of 370 ppm, slightly higher than the earlier reported bulk value of 0.03% $K_2O$ or 250 ppm K (Janots et al. 2012). The soil samples yielded activities corresponding to 0.58 and 0.70 wt.%K.

### Activities of $^{26}$Al in unpaired ordinary chondrites

$^{26}Al$ activities of 30 measured chondrite samples average 52.8±9.1 dpm/kg (range 38.6-79.2, median 51.1). Table 4 and Figure 3 show our results in comparison with values from other surveys. Our data compare well with the mean of 233 non-Antarctic meteorites reported by Nishiizumi (1987) or the complete list of Antarctic meteorites in US collections. Generally low $^{26}Al$ values indicating a high terrestrial age, are reported by Welten (1999) for Lewis Cliff ordinary chondrites and to some degree by Evans (1992) for Victoria Land meteorites. Among our samples, the group of "old" meteorites (50.5 ± 8.0, median 47.4 dpm/kg) is not significantly depleted in $^{26}Al$ as compared with the group of young meteorites (51.1 ± 12.5, median 50.8 dpm/kg). Two unweathered (young) samples show low values of 39-40 dpm/kg, indicating very small preatmospheric masses. Without these two, the young group yields 57 ± 11, median 52.7 dpm/kg. A depletion from 57 to 51 dpm/kg would correspond to an age of 110 kyr but is not considered significant.

### Paired samples from the JaH 091 strewn field

The significant variability of $^{26}Al$ (47.4 to 61.8 dpm/kg, weighted mean 52.0 ± 3.6 dpm/kg) for samples from the large JaH 091 strewn field is expected in case of large meteoroids where the production rate is both dependent on the depth (shielding) and the total size of the meteoroid (Leya and Masarik 2009). Figure 4A shows production rate profiles for L chondrites with different radii, together with the measured activities. Based on the model of Leya and Masarik (2009), we have calculated the volume-averaged bulk mean $^{26}$Al activities of large spherical L chondrite meteoroids (Figure 4B). Based on this model and assuming representative sampling of the whole body, our mean activity of 52.0 ± 3.6 dpm/kg corresponds to a radius of 115 ± 15 cm or a mass of 21 t (range 13.8-30 t, density 3.3 t/m$^3$), in good agreement with the estimate of 9-23 t based on the recovered mass of 4.5 t and an assumed 50-80% mass loss during atmospheric transit. The highest measured activity of 61.8 ± 1.2 dpm/kg, measured in a subsample of the largest mass, is slightly higher than the highest calculated value for a body with a radius of 1 m and possibly indicates an irregular shape. No systematic trend of $^{26}$Al activities versus meteorite mass or position in the strewn field can be deduced (Figure 5), indicating that the distribution of meteorites in the strewn field is not well correlated with the original position in the meteoroid.

**Activities of $^{60}$Co in ordinary chondrites**

$^{60}$Co was not detected in any of the analyzed meteorites (see Table 1). Its production rate strongly increases with shielding to depths of several tens of cm because it is produced from thermal neutrons (Wieler et al. 1996). With a half life of 5.3 years, activities as high as 300 dpm/kg in large meteoroids and a detection limit of ~1 dpm/kg this isotope would be detectable in samples with terrestrial ages up to ~40 years.

**The search for recent falls using $^{22}$Na**

$^{22}$Na ($t_{1/2}$ 2.604 yr) was detected in a single meteorite of the “low terrestrial age group” (as based on $^{14}$C dating and lowest degree of weathering). SaU 424 showed a $^{22}$Na activity of 7.4±0.9 dpm/kg, corresponding to a terrestrial age of 6-9 years before measurement (2007) assuming an initial $^{22}$Na/$^{26}$Al ratio between 1.0 and 1.8 (Bhandari et al. 2002). The interpretation of SaU 424 as a recent fall is consistent with the low activity of $^{26}$Al (39.9±2.2 dpm/kg) and a low mass of 23.2 g, indicating a low pre-atmospheric mass. This also explains the low $^{60}$Co activity (<0.7 dpm/kg).

Sample RaW 034 showed possible detection of $^{22}$Na at the 2 sigma level, but not at the 3 sigma level. The relatively high activity of $^{26}$Al of this sample (73.2±4.8 dpm/kg) requires significant shielding (e.g. at the center of a 35 cm radius meteoroid). The $^{14}$C activity of this sample (23.9±1.0 dpm/kg, Zurfluh et al. 2016) is inconsistent with a recent age and such a shielding, however, but indicates a terrestrial age of 6.9±1.3 kyr (assuming a saturation activity of 55.2 dpm/kg) or 9.7±1.3 kyr for a saturation of 77 dpm/kg (at center of 35 cm meteoroid). Also, the shielding indicated by $^{26}$Al should result in detectable $^{60}$Co for a recent fall. We thus exclude a recent age for this meteorite.

Because SaU 424 is one of the least weathered among a large collection of samples, and we have analyzed all meteorites of our collection with a similarly low degree of weathering, we are confident to have identified all recent falls in our collection. A bias due to preferred sampling of fresh falls by other collectors in the areas visited by us later on would require an extremely selective sampling by others and appears highly unlikely. Assuming a maximum detectable terrestrial age of 20 years using $^{22}$Na and an estimated surface area of 420 km$^2$ searched for meteorites during the Omani-Swiss search campaigns up to 2006 (only from these campaigns all fresh samples were counted), a minimum fall rate of ~120 events per million square km and year is obtained. Based on the fall rate of ~80 events/10$^6$km$^2$*year (Halliday et al. 1989) 0.84 meteorites can be expected in the searched area. We conclude that the number of recent falls is consistent with the fall rate of Halliday et al. (1989).

**Individual meteorites**

In addition to ordinary chondrites, we analyzed Martian shergottite SaU 094 (Gnos et al., 2002), the IIIAB iron Shişr 043 (Al-Kathiri et al., 2006) and howardite JaH 556 (Janots et al. 2012).

*Mars meteorite SaU 094:* SaU 094 yields 152 ± 5 ppm K and 19.5 ± 1.4 ppb Th, in good agreement with values for paired SaU 005 of 0.022% $K_2O$ (=183 ppm K) and 12 ppb Th (Dreibus 2000). The value of 47 ppb $U_{eq}$ indicates recent U series nuclide accumulation. Our $^{26}$Al activity for SaU 094 of 39.1 ± 1.3 dpm/kg is consistent with data obtained on paired samples: 37.4 dpm/kg (Pätsch et al. 2000), 39.5 ± 0.9 dpm/kg (Nishiizumi et al. 2001) and 48.5 ± 1.7 dpm/kg (Bastien et al. 2003). With a saturation activity of 50-80 dpm/kg $^{26}$Al for shergottites having a radius of up to 20 cm (Leya and Masarik 2009), the measured activity corresponds to a degree of saturation of ~50-80%, consistent with an ejection age of 1.2 Ma and a corresponding saturation level of 70% (Mohapatra et al. 2009).
*Iron meteorite Shişr 043*: The $^{26}$Al activity of 3.8 ± 0.3 dpm/kg is close to a value obtained on a different subsample by AMS (3.94 ± 0.12 dpm/kg, Al Kathiri et al. 2006). This value is consistent with a small pre-atmospheric size (radius of 15 cm), as indicated by several other cosmogenic nuclides and noble gas data (Al Kathiri et al. 2006).
*Howardite JaH 556:* This is a mixed HED-chondrite impact melt breccia with large vesicles visible on the surface (Janots et al. 2012). Due to the unusual appearance of this sample we analyzed it with GRS prior to cutting to verify the meteoritic nature. The presence of $^{26}$Al confirmed that it is a meteorite. The $^{26}$Al activity of howardite JaH 556 is 40.9±2.4 dpm/kg, lower than the saturation value on the surface of a meteoroid with 10 cm radius (51 dpm/kg), indicating a small pre-atmospheric size. All larger meteoroids with radii up to 1 m result in higher activities. Alternatively, the sample could have a terrestrial age of at least 230 kyr. Howardites typically have CRE ages of several Ma (Eugster et al. 2006), indicating that $^{26}$Al was saturated at fall time. A high terrestrial age would be consistent with the high degree of weathering and an elevated U/Th ratio of 1.1 and strongly elevated Sr, Ba concentrations (Janots et al. 2012). JaH 556 yielded 0.22 ppm Th and 0.24 ppm $U_{eq}$ ($Th/U_{eq}$ = 0.92), indicating significant terrestrial U-series isotope accumulation.

**Contamination of ordinary chondrites** with **$^{238}$U-/$^{232}$Th-series nuclides and $^{137}$Cs**

The measured activities of $^{232}Th_{eq}$ are relatively uniform for all chondrites and correspond to an average of 50 ± 12 ppb Th (mean L chondrites: 40 ± 7 ppb, Th/U = 3.53, Goreva and Burnett 2001). Good agreement is seen between the values from the different transitions, before and after $^{220}$Rn in the decay chain. $^{238}U_{eq}$(214) averages 32 ± 23 ppb (range 10.8-103, median 24.4). Natural U concentrations are in the range of 10-15 ppb for L chondrites. Again good agreement is seen between the activities from the different transitions used to compute $^{238}U_{eq}$(214). However most activities $^{238}U_{eq}$(226) are higher than those of $^{238}U_{eq}$(214) (mean $^{238}U_{eq}$(226)/$^{238}U_{eq}$(214) activity ratio 1.52 ± 0.20). In 2 samples, however (0603-235 and Al Huqf 010), the ratio $^{238}U_{eq}$(226)/ $^{238}U_{eq}$(214) is clearly smaller than 1. The higher values can be explained by the special chemical affinity of Ra, by diffusion from the Omani soil. $^{232}Th_{eq}$ and $^{238}U_{eq}$ of two soil samples are in the range of typical terrestrial values with $Th_{eq}/U_{eq}$ of 2.8-3.0. A plot of $^{232}Th_{eq}$ versus $^{238}U_{eq}$ of the measured chondrites is shown in Figure 6. Note the trend for systematically elevated $^{238}U_{eq}$ concentrations as compared with $^{232}Th_{eq}$.

Contamination of the meteorites with uranium in Oman meteorites has been documented repeatedly (Stenger et al. 2006, Al-Kathiri et al. 2005, Janots et al. 2012). Contamination is expected, because most analyzed samples have a significant naturally weathered surface area (except Al Huqf 010 cube which has no natural surface and the lowest activities of contaminant nuclides). $^{238}$U series isotopes are likely taken up from soil pore-water, similar as in the case of Sr and Ba (Al-Kathiri et al. 2005, Zurfluh et al. 2012). Excess $^{238}U_{eq}$ averages 33 ppb for the old age group and 9 ppb for the young age group. Even though isotopes of uranium ($^{238}$U, $^{235}$U, $^{234}$U) likely are also taken up, they are not detectable by GRS in meteorites with a typical terrestrial age of 25 kyr, because the daughter elements that are detectable have not grown in significantly in this time. Unsupported $^{226}$Ra ($t_{1/2}$ 1602 years) decays within a fraction of the typical terrestrial life of many Omani meteorites. Systematic contamination of Omani meteorites with barium (Al-Kathiri et al., 2005) is consistent with the observed enrichment of the geochemically similar element radium. $^{226}$Ra uptake or $^{222}$Rn loss is also indicated by a systematically higher activity of $^{226}$Ra as compared with $^{238}U_{eq}$ which is measured from combined peaks of $^{214}$Bi and $^{214}$Pb further down in the decay chain (Figure 7). We conclude that $^{226}$Ra is accumulated in the meteorites during weathering, and that this activity is partly lost again, most likely due to escape of $^{222}$Rn.

The situation is different for $^{232}$Th, which is highly insoluble and has no daughter elements of geologically significant lifetime. For this element, contamination likely is dominated by particulate accumulation.

The contamination with man-made $^{137}$Cs is variable but generally low. It was detected in 27 out of 33 meteorite samples, with a median of 1.1 and a maximum of 10.3±1.6 dpm/kg. Komura et al. (1982) have analyzed Antarctic meteorites for $^{137}$Cs contamination, reported as net counts only. Their mass-normalized (not further corrected) data correspond to a median of 1.3 and a maximum of 3.0 dpm/kg, similar to the range of our data. The soil samples show much higher activities (34.0 ± 0.8 and 210.0 ± 1.8 dpm/kg), but still much lower than in typical European soils (1-100*$10^3$ dpm/kg, Pourcelot et al. 2007). $^{137}$Cs activities are weakly correlated with $^{232}$Th (r = 0.49). This is probably due to tightly adhering soil (enriched in $^{232}$Th) that also adsorbs $^{137}$Cs. Several meteorites have $^{137}$Cs/$^{232}$Th activity ratios significantly higher than the soils, indicating adsorption of $^{137}$Cs onto the meteorites from solution, and not just particulate contamination.

## CONCLUSIONS

GRS data confirm the absence of very old (>300 kyr) ordinary chondrites among the Omani meteorite population. We have identified a single recent fall on a the searched surface of ~420 km$^2$, consistent with estimated fall rates (Halliday et al., 1989). The mean activity and range of activities of $^{26}$Al of 12 samples from the 52 km-long JaH 091 strewn field indicates a meteoroid radius of 115 ± 15 cm.

Omani meteorites show contaminations with natural and man-made terrestrial radioisotopes, in particular recent uptake of $^{226}$Ra, consistent with the observed long-term accumulation of Ba in hot desert meteorites.

This study demonstrates the suitability of GRS for determining cosmogenic isotopes ($^{26}$Al and potentially others), primordial radioelements ($^{232}$Th, $^{40}$K) and man-made contaminants ($^{137}$Cs) in hot desert meteorites. Due to the mobility of several isotopes in the $^{238}$U decay chain, the values obtained for $^{238}U_e$ and $^{226}$Ra are likely affected by disequilibria and/or inhomogeneous distribution within the meteorites, so that these values must rather be taken as indicators for recent disturbances. Given the potential for strong regional variations in natural (U, Th-series) and artificial ($^{137}$Cs) radioactive contaminants, these may be useful for source-fingerprinting of meteorites when suitable reference samples are available.

In particular, we demonstrate the potential to identify recent falls among meteorites collected in hot deserts by analyzing a selection of the least weathered samples using gamma-ray spectroscopy. We see a potential for future analyses of hot desert meteorites in order to identify very young falls (using $^{22}$Na, $^{60}$Co and potentially $^{44}$Ti) as well as very old ones using $^{26}$Al. For unusual meteorites, the ability to obtain bulk Th, (U) and K concentrations non-destructively on large samples is a valuable tool for characterization and even source determination. We plan further analyses of hot desert meteorites with the aim to cover all our finds of fresh meteorites using a new gamma ray spectrometer operational at the Vue des Alpes tunnel since December 2015 (von Sivers et al. 2015).

*Acknowledgements* - Meteorite search campaigns were conducted with kind permission from the Directorate General of Minerals, Ministry of Commerce and Industry, Muscat, Sultanate of Oman. We particularly thank director general of minerals, Salim Al-Ibrahim, Mohammed Al-Rajhi and Mohammed Al-Battashi. We thank all participants in the field campaigns. The manuscript was significantly improved based on comments by Kees Welten and an anonymous reviewer.

**Tables:**

Table 1: Radioisotope activity data for all measured samples, the errors are purely statistical. A 10% additional systematic uncertainty is estimated.

Table 2: Activities of radioisotopes measured on a naturally shaped sample and on a cut parallelepiped sample of a large single individual of the Al Huqf 010 (L6) meteorite.

Table 3: Comparison of GRS data obtained at Vue-des-Alpes with data from the Modane laboratory and an AMS determination at the University of California, Berkeley.

Table 4: Comparison of $^{26}Al$ activities (dpm/kg) of ordinary chondrites from Oman with Antarctic and non-Antarctic populations.

**Figure captions:**

Figure 1: Typical γ-ray spectrum of the Ge detector at Vue-des-Alpes tunnel. Background spectrum (no sample) and meteorite sample 0603-235 (JaH 091 pairing).

Figure 2: Detection efficiency for a typical meteorite; circles: single gamma, cross: taking into account pile ups ($^{208}Tl$, $^{214}Bi$, $^{26}Al$, $^{22}Na$, $^{60}Co$)

Figure 3: Boxplot for $^{26}Al$ activities of different populations of ordinary chondrites. Shown are median, boxes containing 50% of the data and whiskers of 1.5x box width. Note the similarity of all population including Oman, with the exeption of Lewis Cliff, Antarctica. Data source: 1, This study; 2, Welten (1999); 3, Evans (1992); 4, U.S. Antarctic collection: http://curator.jsc.nasa.gov/antmet/us_clctn.cfm; 5, Nishiizumi (1987).

Figure 4: Estimation of meteoroid radius from the $^{26}Al$ activity in 12 samples of the JaH 091 strewn field. A) Activities of $^{26}Al$ as a function of position in L chondrite meteoroids with radii of 85 to 150 cm based on the model of Leya and Masarik (2009). Also shown are activities measured in JaH 091, and their weighted mean (wm). B) Volume-averaged mean $^{26}Al$ activities for spherical L chondrites with radii of 10 to 200 cm. The weighted mean of $^{26}Al$ in JaH 091 and pairings is 52.0±3.6 dpm/kg. This value yields two size solutions (r~20 cm and r=115±15 cm), the lower one inconsistent with the large mass recovered.

Figure 5: Activities of $^{26}Al$ in different individuals of the ~52 km JaH 091 strewn field. The position in the strewn field is given as distance from the largest mass (~1400 kg). The larger the distance, the smaller the masses. See Table 1 for individual masses.

Figure 6: Plot of concentrations of $^{238}U_{eq}$ versus $^{232}Th$. Note the generally increased $U_{eq}$/Th ratio as compared with mean chondrites, indicating accumulation of $^{238}U$ daughter elements.

Figure 7: Plot of the activities of $^{238}U_{eq}(226)$ versus $^{238}U_{eq}(214)$ in function of meteorite mass.